\documentclass[aps, 11pt, superscriptaddress]{revtex4-2}
\usepackage{amsfonts}
\usepackage{amssymb}
\usepackage{amsmath}
\usepackage{booktabs}
\usepackage{graphicx}
\usepackage{dcolumn}
\usepackage{bm}
\usepackage{units}
\usepackage{multirow}
\usepackage{CJKutf8}
\usepackage{subfigure}
\usepackage{color}
\usepackage{xcolor}
\usepackage{url}
\usepackage[colorlinks,linkcolor=blue,anchorcolor=blue,citecolor=blue,urlcolor=blue]{hyperref}
\usepackage[utf8]{inputenc}
\DeclareUnicodeCharacter{03C6}{\ensuremath{\phi}} 

\begin{document}

\author{Hechen Ren}
\email[Corresponding author: ]{ren@tju.edu.cn}
\affiliation{The International Joint Institute of Tianjin University, Fuzhou, Tianjin University, Tianjin 300072, China}
\affiliation{Center for Joint Quantum Studies \& Tianjin Key Laboratory of Low Dimensional Materials Physics and Preparing Technology, Department of Physics, School of Science, Tianjin University, Tianjin 300072, China}

\author{Ziying Li}
\affiliation{The International Joint Institute of Tianjin University, Fuzhou, Tianjin University, Tianjin 300072, China}

\title{The Superconducting Talbot Effect in Phased-Array Josephson Junctions}

\begin{abstract}
We introduce the superconducting Talbot effect---a macroscopic quantum self-imaging phenomenon occurring when proximitized Cooper pairs propagate through a ballistic two-dimensional electron gas. By configuring periodic superconducting leads into a phased-array Josephson junction with programmable phase differences, we demonstrate active steering of the resulting superconducting Talbot carpet. To overcome transport resolution limits, we design a Vernier-scale collector array that performs sub-wavelength sampling of the fractional Talbot pattern. This approach maps real-space quantum interference with high robustness to disorder, enabling direct extraction of Fermi wavelengths across helical, spin-degenerate, and spin-orbit-split Fermi surfaces. Tight-binding numerical calculations on a square lattice validate the real-space interference patterns. Our results establish a versatile framework for coherent wavefront engineering and quantum materials diagnostics in superconducting optics.
\end{abstract}

\maketitle
\newpage

\section*{Introduction}

When coherent waves pass through a periodic diffraction grating, they undergo lateral interference that replicates the spatial profile of the grating at regular longitudinal intervals. This near-field self-imaging phenomenon is known in classical optics as the Talbot effect \cite{Talbot1836, Case2009}. Its principles have transcended electromagnetism to manifest in diverse physical systems spanning acoustic waves, surface plasmons, and atoms \cite{Wen2013, Chapman1995}. Lately, the paradigm has theoretically extended to solid-state systems where the de Broglie waves of ballistic electrons in two-dimensional electron gases (2DEGs) such as graphene undergo self-imaging when modulated by static potential gratings \cite{Salas2016, Walls2016}. Yet, the Talbot effect has remained unstudied in the context of superconductivity---a state known for its macroscopic phase coherence \cite{Devoret1985, Martinis1985}.

Planar multi-terminal Josephson junctions (MTJJs) \cite{Pankratova2020, Chiles2023, Gupta2023a, Lesser2024, Zhu2025} offer a premium playground to realize this macroscopic quantum manifestation of near-field diffraction. If we control the terminals' superconducting phases externally, the device acts as a phased-array Josephson junction (PAJJ). This architecture borrows from radar engineering, where programmable phase delays between adjacent antennas can steer the beam in the far field \cite{Parker2002} and the near field \cite{Li2019c}. Yet, this rich wavefield physics has remained unexplored in superconductivity. Current measurements on MTJJs remain spatially averaged \cite{Tinkham2004}, washing out the fine spatial variations of the Andreev states. Consequently, traditional macroscopic measurements are largely blind to the local landscape of the induced order parameter and the size and texture of material's Fermi surface. Resolving the Fermiology in the proximitized material would unveil a complete picture of how coherent electron-hole pairs form Andreev bound states and propagate in space. It would also expand our tools to characterize Fermi wavelengths in 2DEGs, which are currently limited to magnetoresistance oscillations, scanning tunneling spectroscopy, and angle-resolved photoemission spectroscopy \cite{Ando1982, Crommie1993, Damascelli2003}.

In this work, we introduce the superconducting Talbot effect and establish the PAJJ as a novel platform to study such light-like propagation of induced superconductivity. We show that tuning the junction material's Fermi level produces oscillations in the critical current. To measure arbitrary wavelengths, we combine active phase control in a PAJJ with Vernier-scale collectors that perform sub-wavelength spatial sampling. We evaluate the model across distinct physical regimes: a single helical Fermi surface, spin-degenerate Fermi surfaces, and Fermi surfaces split by spin-orbit coupling. In the helical case, we steer the Talbot carpet and directly extract the wavelength from the lateral movement. In the spinful cases, we provide a regressive method to fit the Fermiology from the current-phase relations across the collector array. Finally, we verify our results using tight-binding numeric simulations on a square lattice. Our work opens new avenues for quantum materials diagnostics and ushers in a new era of coherent wavefront engineering in superconducting optics.

\par

\section*{Results and Discussion}

\subsection*{The Superconducting Talbot Effect}
We consider an array of $N$ superconducting source leads located at $y=0$, from which superconductivity is proximitized into a 2DEG occupying the $y > 0$ half-plane (Fig. 1a, b). Each of the source leads has width $s$ and the period of the array, namely the pitch, is $d$. For illustration purposes, we choose $N=20$, $s = 100$ nm, $d = 200$ nm. To see the emergence of the superconducting Talbot carpet in real space, we start from the propagator $F(x, x_s, y)$ for an induced Cooper pair from point $(x_s, 0)$ in a source lead to point $(x, y)$ in the 2DEG \cite{Hart2016}. The induced superconducting order parameter is thus

\begin{equation}
\Delta(x,y) = \sum_{m=0}^{N-1} \int_{md}^{md+s} dx_s \Delta(x_s, 0) F(x, x_s, y)
\end{equation}

where $\Delta(x_s, 0)$ is the complex superconducting coupling at source position $(x_s, 0)$. 

\begin{figure}[htbp!]
\centering
\includegraphics[width=0.75\linewidth]{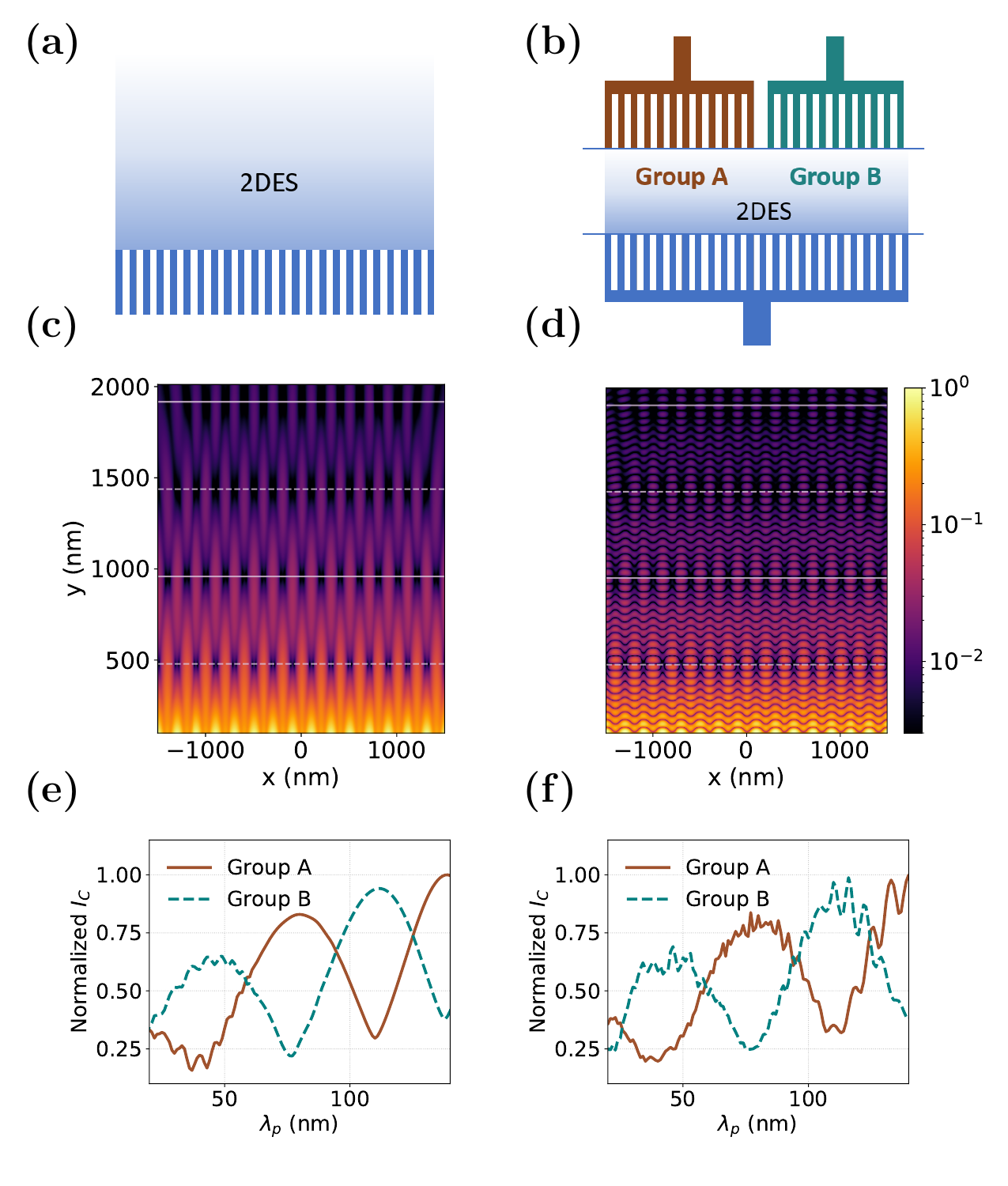}  
\caption{
\textbf{The superconducting Talbot effect.}
(a) Schematic of superconductivity induced in a 2DEG by a source array. 
(b) Measurement scheme illustrating the two groups of collector array, where Group A aligns while Group B misaligns with the source. 
(c,d) Real-space intensity of induced superconductivity for a single helical (c) and spin-degenerate (d) Fermi surfaces. Solid (dashed) lines mark integer (half integer) multiples of the Talbot distance.
(e,f) Total critical current $I_c$ between the source and each collector group versus the pair wavelength for a single helical (e) and spin-degenerate (f) Fermi surfaces.}
\label{fig_1}
\end{figure}

The simplest case is a 2DEG with a single helical Fermi surface and a Fermi wavenumber $k_F$, such as the surface of a 3D topological insulator \cite{Chen2009a}. The propagator

\begin{equation}
F(x, x_s, y)\propto \frac{e^{i 2k_F r}}{r^2},
\label{single_FS_GF}
\end{equation}

where $r = \sqrt{y^2 + (x-x_s)^2}$ \cite{Hart2016, Zhang2024c}. The effective wavelength for the Cooper pair is $\lambda_p = 2\pi / 2k_F = \pi/k_F$. Practically, we assume $\lambda_p$ is smaller than the pitch $d$ but not by orders of magnitude. The resulting induced superconducting order parameter is shown in Fig. 1c, rendering a Talbot carpet in the $y > 0$ half-plane. The images of the source array at the bottom of the panel are repeated at integer multiples of the Talbot distance \cite{Rayleigh1881}

\begin{equation}
Z_T = \frac{\lambda_p}{1 - \sqrt{1 - \frac{\lambda^2_p}{d^2}}}
\label{double_FS_GF}
\end{equation}

which simplifies to $ \frac{2d^2}{\lambda_p}$ in the limit $ \lambda_p \ll d $. At half integers, we obtain inverted images of the source array, where bright spots become dark spots but the period is maintained. Here, the pair wavelength is assumed to be $\lambda_p = 80$ nm, which corresponds to a Talbot distance of $Z_T \approx 958$ nm.

In the case of a spin-degenerate Fermi surface in the 2DEG, we have two components in the pair wavefunction. The propagator 

\begin{equation}
F(x, x_s, y)\propto \frac{\cos(2k_F r) }{r^2}.
\end{equation}

The Talbot carpet becomes more textured due to the standing-wave interference (Fig. 1d), yet its overall envelope mimics that of the single-component case (Fig. 1c) with the signature revivals and reversals of the Talbot effect (see Supplementary Figure S1 for linecuts of Fig. 1c,d).

To measure this effect, we place collector leads and measure critical current of the resultant Josephson junction, which is proportional to the induced superconducting gap at the collector site \cite{Hart2016}. The collector array at $y=L$ features the same lead width and spacing (Fig. 1b). To monitor the shift in the Talbot self-image, we can arrange Group A of the collectors to be aligned to the source array while keeping Group B collector leads aligned to the gap of the source leads (perfectly misaligned). Each group can be connected via a normal bond pad at the end to be measured collectively for a better signal-to-noise ratio. In both types of Fermi surfaces, the critical currents feature prominent oscillations as the wavelength varies (Fig. 1e,f) with the collectors at a fixed distance $y = 1 \; \mu$m (see Supplementary Figure S2 for different $y$-distances). The critical current of Group A reaches its maximum when the junction length $L$ matches the full Talbot distance, while Group B reaches its maximum at half the Talbot distance. Due to the fast spatial oscillations in the spin-degenerate case, data in Fig. 1f has been 10-point smoothed (raw data in Supplementary Figure S3).

\subsection*{The Steerable Talbot Carpet}
In addition to the integer self-images, partial recurrences also take place at fractions of the Talbot distance, known as the fractional Talbot effect \cite{Berry1996}. This liberates us from the resonance conditions when studying the Fermiology of a 2DEG. Given the induced superconductivity decays with $r^2$, planar Josephson junctions typically have a length between the $100$ nm and $1 \; \mu$m. Therefore, the collector position coincides with a fraction of the Talbot distance for a large range of realistic Fermi wavelengths (between $10$ and $100$ nm).

\begin{figure}[htbp!]
\centering
\includegraphics[width=0.75\linewidth]{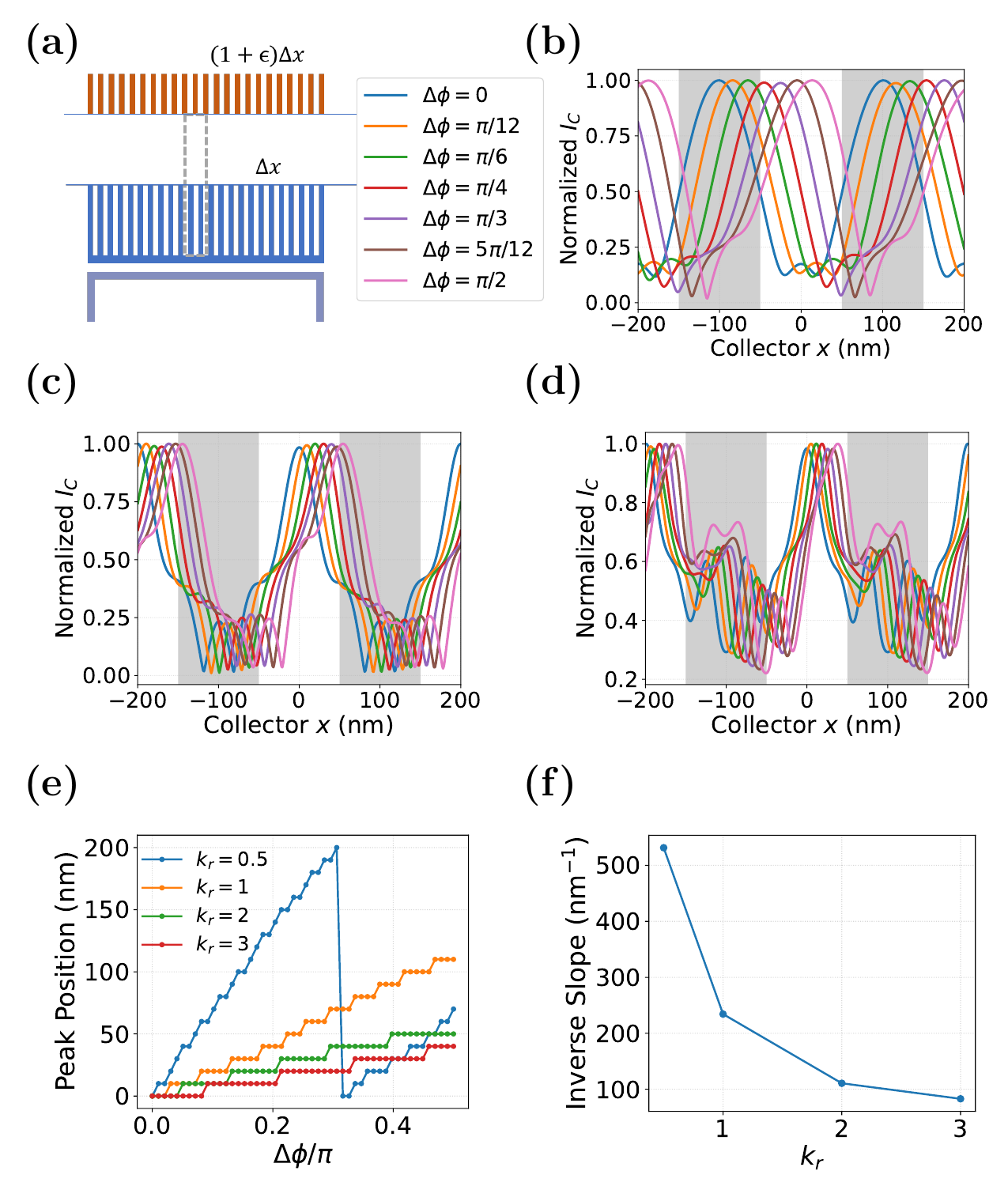}  
\caption{
\textbf{Steering the fractional Talbot fringes for wavelength measurement.}
(a) Schematic of a PAJJ with Vernier scale for the collector array. (b-d) The fractional Talbot interference fringes over two spatial periods at the collector array (gray dashed box in (a)) for increasing Fermi wavenumbers $ k_p = \frac{\pi y}{d^2} $ (b), $\frac{2\pi y}{d^2}$ (c), and $\frac{3\pi y}{d^2}$ (d). The corresponding phase differences $\Delta \phi$ are labeled to the left of (b). (e) Peak position as a function of $\Delta \phi$ for different $ k_r = \frac{d^2}{\pi y} k_p$. (f) Extracted slope for each curve in (e).}
\label{fig_2}
\end{figure}

To measure arbitrary Fermi wavelengths using the superconducting Talbot effect, we replace the uniform-phased source array with a 1D PAJJ along $x$ (Fig. 2a). By running a mA-level current through the thistle-colored lead at the bottom, we can flux-bias the phase loops. This generates a constant phase difference $\Delta \phi$ between each pair of adjacent source leads and results in a linear phase profile in the PAJJ $\phi_m = m \Delta \phi$.

In the case of one helical Fermi surface, the induced order parameter in the 2DEG region becomes

\begin{equation}
\Delta(x,y) \propto  \sum_{m=0}^{N-1} \int_{md}^{md+s} dx_s \frac{e^{i \phi_m}e^{2i k_F r }}{r^2}.
\end{equation}

This effectively injects the Cooper pairs with a momentum \cite{Ye2026} and results in a steering effect of the outgoing wave, analogous to the classical phased-array antennas \cite{Parker2002}. With a phase delay $\Delta \phi$, the steer angle is thus given by $\theta = \mathrm{arcsin} \left( \frac{\lambda_p}{2 \pi d} \Delta \phi \right)$ \cite{Balanis2016}, and the lateral shift of the fractional Talbot pattern at a fixed distance $L$ becomes

\begin{equation}
\Delta x (L) = \mathrm{arcsin} \left(\frac{\lambda_p}{2 \pi d} \Delta \phi\right) L,
\end{equation}

which directly scales with the Fermi wavelength. Therefore, if we could monitor the lateral crawl of the Talbot pattern as a function of the source array's phase difference, we could extract the Fermi wavelength of the underlying 2DEG.

Since this steered crawl could be sub-pitch-scale, we need to discern the fineprints of the fractional Talbot effect. If the collector array perfectly matches the source, it becomes blind to the high-frequency spatial variations of the Talbot pattern. To better resolve the Talbot fringes, we design a Vernier-scale collector array with collector pitch $d_c = d(1 + 1/N)$ (Fig. 2a). The first collector is aligned with the source probe, and each subsequent collector is shifted by $d/N$, sampling a slightly different spot within the lateral period. This Vernier array effectively takes $N$ independent spatial slices of a single unit cell in the Talbot pattern, boosting effective spatial resolution by a factor of $N$ without pushing the lithographic limit.

As shown in Fig. 2b-d, the higher (lower) the Fermi wavenumber (wavelength), the smaller the lateral shift at the collector location $y = 1 \; \mu$m. Tracking the peak position of the fractional Talbot fringes gives us a direct measurement of the Fermi wavelength in this single-component case. Considering we have $N = 20$ discrete critical-current measurements, the peak position will exhibit staircase movements (Fig. 2e). The smaller the pair wavelength, the faster the movement as a function of $\Delta \phi$, resulting in multiple periods within the $\pi/2$ steer window in the fastest case. Figure 2f shows the extracted slope versus the wavelength, displaying an approximate $1/x$ correlation. The full Talbot carpet under steering is shown in Supplementary Figure  S4.

\subsection*{Full Fermiology} 
When the 2DEG possesses two spin-degenerate or spin-split Fermi surfaces, the resulting Talbot fringes remain static like a standing wave without apparent lateral shift. Study of the full Fermiology requires us to examine the critical currents at all the collector probes. Considering the probes are discrete while the source phase difference $\Delta \phi$ can be continuously swept, this collection of $N-1$ functions of $\Delta \phi$ over the full $2\pi$ period form the unique fingerprint of the Fermi surfaces with their spin textures. Figure 3 shows the sets of critical currents for spinless (Fig. 3a-c), spin-degenerate (Fig. 3d-f), and a 10\%-spin-split Fermi surfaces with spin-orbit coupling (Fig. 3g-i).

\begin{figure}[htbp!]
\centering
\includegraphics[width=0.75\linewidth]{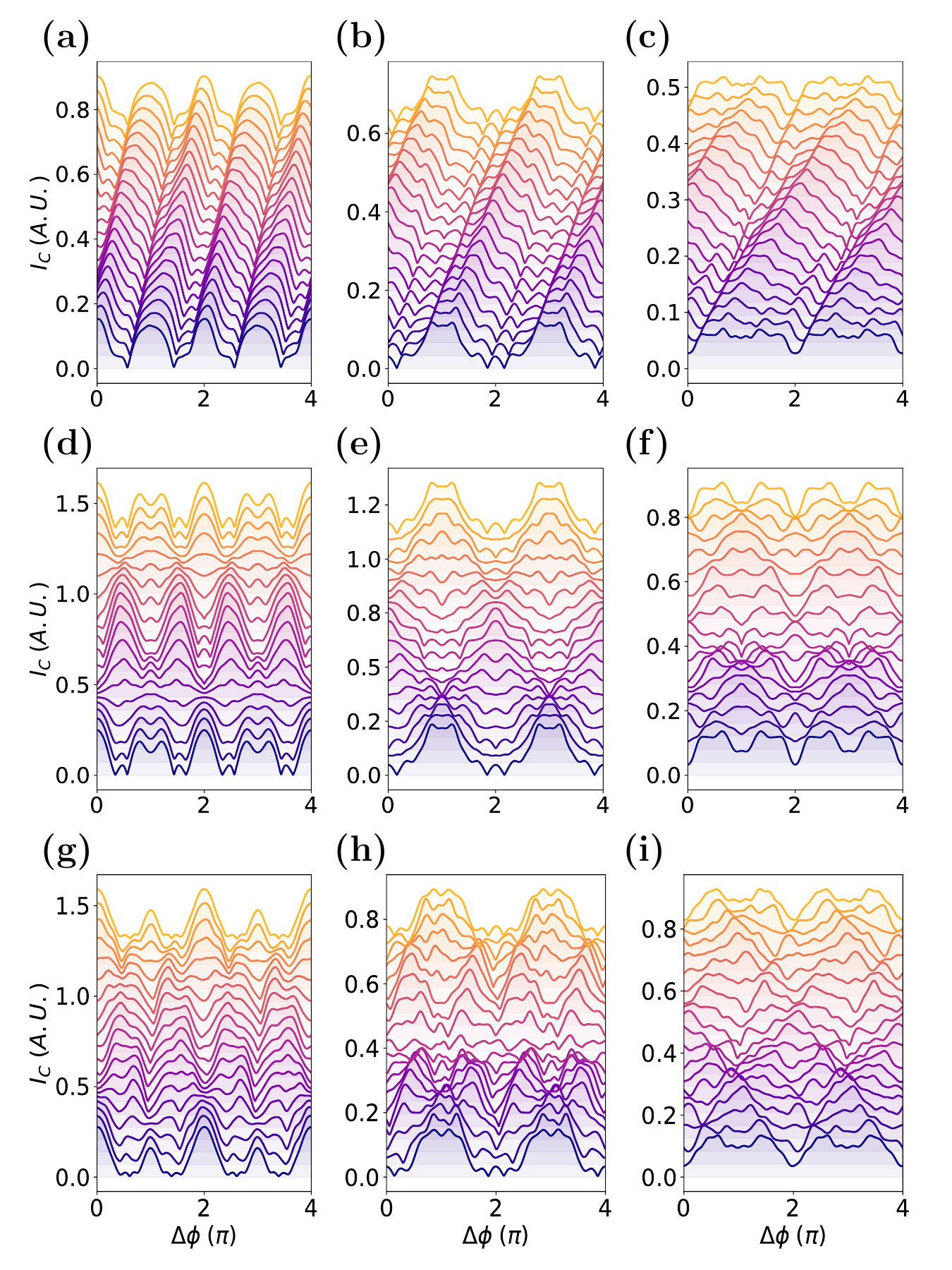}  
\caption{
\textbf{Phase-dependent critical currents for full Fermiology.}
Critical currents between the source array and each of Vernier-scale collectors (colored blue to orange) as functions of the source phase difference $\Delta \phi$ over two periods. The pair wavelength increases from $\frac{\pi y}{d^2}$ to $2\frac{\pi y}{d^2}$ to $3\frac{\pi y}{d^2}$ from left to right for single helical (a-c), spin-degenerate (d-f), and 10\%-spin-split (g-i) Fermi surfaces, where $y = 1 \; \mu$m and $d = 200$ nm.}
\label{fig_3}
\end{figure}

As established in the last section, the Talbot carpet for a single helical Fermi surface (Fig. 3a-c) features steerable peaks shifting laterally, and the shifting speed correlates directly with the pair wavelength. In the case of $k_p = \frac{\pi y}{d^2}$ (Fig. 3a), the peak location (which curve possesses the maximal critical current at a given $\Delta \phi$) moves fastest across the collector probes, while it moves slowest in the case of $k_p = 3\frac{\pi y}{d^2}$ (Fig. 3c). However, in the case of the spin-degenerate and spin-split Fermi surfaces, due to the interference between the superconducting components, the peaks split and become harder to track. Therefore, the simple method of plotting peak position versus $\Delta \phi$ loses effect.

Fortunately, even in these complex situations, the full set of $N$ phase-dependent functions corresponding to the Vernier-scale collectors (each panel in Fig. 3) provide enough information to discern the texture of the Fermi surface. Because we can calculate these complete sets of $I_c (\Delta \phi)$ functions for any arbitrary Fermi surfaces (spin-degenerate, spin-split, etc), we could build a regression model to fit the experimental data and diagnose its Fermiology. To show that the fitted solution is unique and accurate, we generate data using numeric simulations for a 2D range of $k_1$ and $k_2$ of a spinful Fermi surface, and we fit them using a regression model via a mean-squared-error cost function. To efficiently navigate the fast oscillations in the cost function, we adopt a two-step fitting protocol. First, we fit the average wavelength of the two spinful Fermi surfaces, $\bar{k} = (k_1 + k_2)/2$. Then, using the resultant $\bar{k}$ as an input, we fit the remaining variable $\Delta k = k_1 - k_2$.  Figure 4a shows the fitted results for various $k_1$ and $k_2$, where most of the obtained values lie on or near the $k_\mathrm{true} = k_\mathrm{fit}$ line. The 2D map of fitting errors are shown in Fig. 4b, exhibiting a maximal error rate of under $9$\% over the entire domain of interest. Similar performances are found if the collectors are placed at closer distances such as $400$ and $200$ nm (Supplementary Figure  S5).

\begin{figure}[htbp!]
\centering
\includegraphics[width=0.75\linewidth]{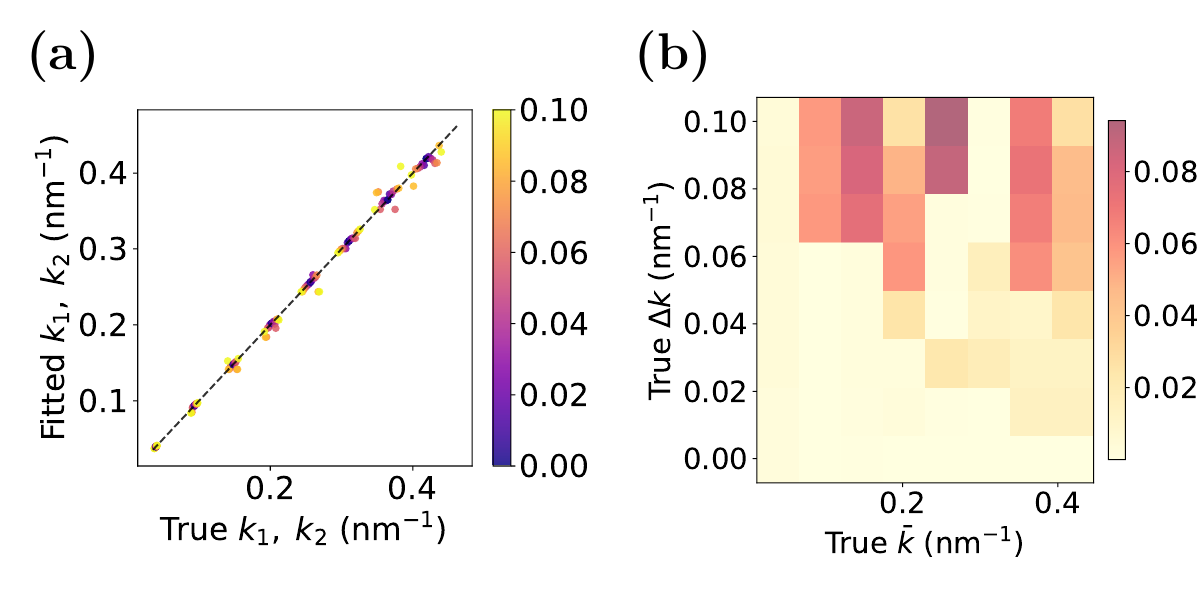}  
\caption{
\textbf{Regression model performance.}
Model accuracy for spinful Fermiology with $y = 1 \; \mu$m and $d = 200$ nm. 
(a) Fitted results for various $k_1$ and $k_2$ combinations. Marker color corresponds to the mismatch $\Delta k / \bar{k}$.
(b) Maximal error rate vs. the average wavelength $\bar{k}$ and mismatch $\Delta k$.}
\label{fig_4}
\end{figure}

\subsection*{Lattice Model}
So far, we have based our theory on analytical formula of the Cooper-pair Green's function (Eqn. \eqref{single_FS_GF} and \eqref{double_FS_GF}). To independently verify the superconducting Talbot effect and demonstrate it in quantum-mechanical systems, we evaluate tight-binding models on a square lattice with nearest-neighbor hopping using the Kwant package \cite{Groth2014}. An array of rectangular region on the bottom of the device represents the superconducting source leads, where the Bogoliubov-de Gennes (BdG) Hamiltonian \cite{DeGennes2018} is given by 

\begin{equation}
H_{\text{BdG}} = \begin{pmatrix} H_0(\mathbf{k}) & \Delta \\ \Delta^* & -H_0^*(-\mathbf{k}) \end{pmatrix},
\end{equation}

with the superconducting order parameter $\Delta = i\Delta_0 \sigma_y$.

For the spin-degenerate case, we use the Hamiltonian for a generic parabolic band \cite{KwantSuperconductorTutorial}.

\begin{equation}
H_0(\mathbf{k}) = 2 t ( 2 - \cos k_x a - \cos k_y a)- \mu ,
\end{equation}

where $t$ is the hopping amplitude, $a$ is the lattice constant, and $\mu$ is the chemical potential. For the single helical Fermi surface, we adopt the Dirac surface-state Hamiltonian \cite{Fu2008, Qi2006, Qi2008}.

\begin{equation}
H_0(\mathbf{k}) = t \sigma_z (2-\cos k_x a - \cos k_y a)
 + t (\sigma_y \sin k_x a  - \sigma_x \sin k_y a ) - \mu \sigma_0.
\end{equation}

Here $\sigma$'s are the Pauli matrices in the spin space. If we stay in the limit $ka \ll 1$, the Hamiltonian simplifies, to first order in $k$, to a Dirac cone with Rashba spin-orbit coupling 

\begin{equation}
H_0(\mathbf{k}) = ta ( k_x \sigma_y - k_y \sigma_x ) - \mu \sigma_0.
\end{equation}

\begin{figure}[htbp!]
\centering
\includegraphics[width=0.75\linewidth]{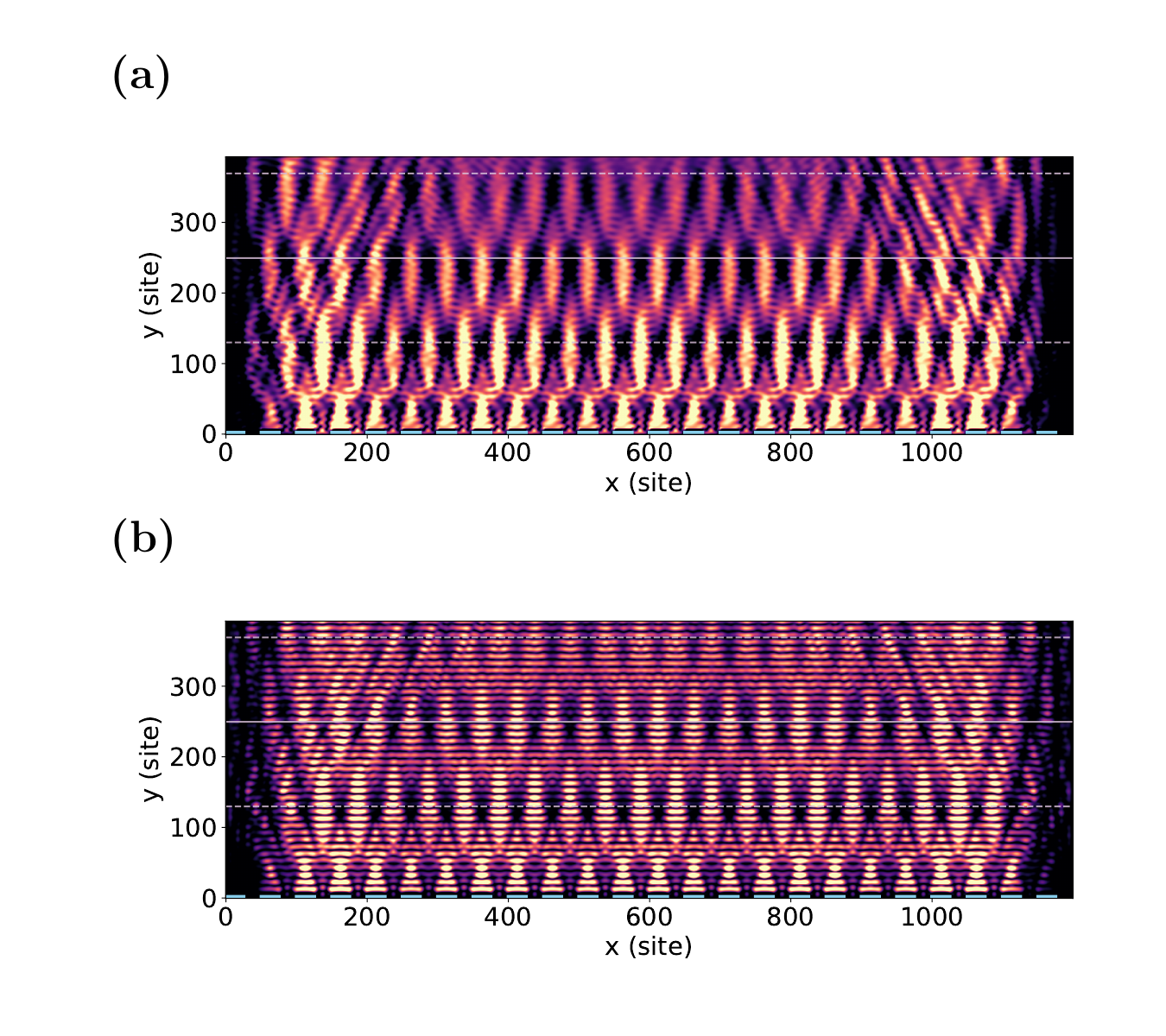}  
\caption{
\textbf{Tight-binding simulation on a square lattice.}
Real-space map of induced superconducting gap with grid spacing $a = 4$ nm, pitch $d = 200$ nm, and pair wavelength $\lambda_p = 80$ nm in the case of a single helical Fermi surface (a) and spin-degenerate Fermi surfaces (b). Solid and dashed lines correspond to the integer and half-integer Talbot distances. The superconducting gap in the source leads is set to be $\Delta = 0.5t$.}
\label{fig_5}
\end{figure}

Figure 5 shows simulation results of the real-space map of the induced superconducting term. The light-blue dashes on the bottom of the panels represent the superconducting source leads, while a normal lead is attached on top of the simulated region. In both cases with a single and double Fermi surfaces, the lattice model renders periodic modulations of the induced superconductivity in two-dimensional space. The source array's self-image appears around the Talbot distance and its negative image around half that distance, reproducing the Talbot carpet we calculated analytically. Supplementary Figure  S6 contains results for different wavelengths.

\section*{Conclusions} 
In summary, we have introduced the superconducting Talbot effect in a PAJJ platform to study the induced superconductivity's light-like propagation. By combining phase-controlled wavefront steering with Vernier-scale sub-wavelength spatial sampling, our architecture provides a quantum-transport probe for mapping Fermi wavelengths and Fermi surface textures without needing external magnetic fields. The gate tunability of a planar structure makes our protocol experimentally feasible and widely applicable to 2D materials. Furthermore, employing Nb or NbN leads can help boost the induced superconducting order parameter and hence the critical currents in individual collectors. One big advantage of the Talbot effect in superconducting devices is its robustness against local disorders, as the Talbot pattern self-heals around scatterers as if blind to them (Supplementary Figure  S7). Our proposal for the Talbot effect in PAJJ establishes an ideal playground and a major step toward the interdisciplinary field of superconducting optics. Moving forward, it would be intersting to tap into independent phase controls as well as the phase information at the collector side and explore digital wavefront shaping analogous to optical engineering \cite{Vellekoop2007, Popoff2010}.
\par\quad\par

\noindent{\large\textbf{Acknowledgements}}\\
We thank the Fundamental and Interdisciplinary Disciplines Breakthrough Plan of the Ministry of Education of China (JYB2025XDXM410). This work is also supported by the National Natural Science Foundation of China.
\par\quad\par

\noindent{\large\textbf{Author contributions}}\\
H. Ren conceived the project, constructed the theory, and prepared the manuscript. H. Ren and Z. Li performed the numerical calculations. 
\par\quad\par

\bibliography{talbot-JJ-theory}

\end{document}